\documentclass[a4paper,11pt]{article}
\usepackage{pos}
\usepackage{framed}
\usepackage{listings}
\usepackage{fancyvrb}
\usepackage[lofdepth,lotdepth]{subfig}
\input{lst-color.sty}
\newcommand{\marty}{\Verb+MARTY+}
\newcommand{\martyv}{\Verb+MARTY-1.4+}
\title{Automatic extraction of one-loop Wilson coefficients in general BSM scenarios using \martyv}
\ShortTitle{Extraction of general Wilson coefficients with \martyv}

\author*[a]{Grégoire Uhlrich}
\author[a,b]{Farvah Mahmoudi}
\author[a,b]{Alexandre Arbey}

\affiliation[a]{Universit\'e de Lyon, Universit\'e Claude Bernard Lyon 1, CNRS/IN2P3, \\
Institut de Physique des 2 Infinis de Lyon, UMR 5822, F-69622, Villeurbanne, France}

\affiliation[b]{Theoretical Physics Department, CERN, CH-1211 Geneva 23, Switzerland}

\emailAdd{marty.uhlrich@gmail.com}

\abstract{We present a fully automated procedure providing an easy way to perform, systematically, phenomenological analyses in flavor physics for general BSM scenarios. This procedure relies on \martyv, is model independent and requires as input only the Lagrangian of the theory. Once the Lagrangian has been defined, tree-level and one-loop Wilson coefficients can be calculated symbolically by \marty, from which flavor observables can be computed numerically by available software programs. We focus in particular on $b\to s\gamma$ and the recently updated $b\to s\ell^+\ell^-$ observables which are in tension with the SM, and present a general procedure to extract the relevant one-loop coefficients, such as $C_7$, $C_7^\prime$, $C_9$ and $C_{10}$.
}

\FullConference{%
  *** The European Physical Society Conference on High Energy Physics (EPS-HEP2021), ***\\
  *** 26-30 July 2021 ***\\
  *** Online conference, jointly organized by Universität Hamburg and the research center DESY ***
}

\begin{document}
\maketitle

\section{Introduction}

In the following we present a sample program using \martyv~\cite{marty} to calculate, in general BSM scenarios, one-loop Wilson coefficients for flavor physics such as $C_7^{(\prime)}$ and $C_{9(10)}$. These coefficients which are relevant to describe the $b\to s$ transitions are defined from the corresponding effective Hamiltonian~\cite{buras}, namely
\begin{equation}
\label{heff}
    \mathcal{H}_{\mathrm{eff}} \ni -\frac{4G_F}{\sqrt{2}}V_{tb}V_{ts}^* \left[\frac{e}{16\pi^2}m_b C_7^{(\prime)}\big(\bar{s}\sigma^{\mu\nu}P_{R(L)}b\big)F_{\mu\nu} + \frac{e^2}{16\pi^2}C_{9(10)}\big(\bar{s}\gamma^\mu P_Lb\big)\big(\bar{\mu}\gamma_\mu(\gamma^5)\mu\big)\right],
\end{equation}
with $b$, $s$ and $\mu$ the SM fermionic fields and $F_{\mu\nu}$ the photon's field strength.

The calculation of these coefficients using \marty\ is model-independent and is not designed specifically for $C_7^{(\prime)}$ or $C_{9(10)}$. As presented in the next section, the procedure is very general and can be used to derive other coefficients, at the tree level or for other types of effective operators. Once these values are calculated by \marty\ symbolically, numerical values of flavor observables can be obtained using \Verb+SuperIso+~\cite{superiso1} for example.

\section{The program}

For more information about \marty\ (installation instructions, manuals, documentation or related publications) see~\url{https://marty.in2p3.fr}.

\paragraph{Model definition}
This part is the only model-dependent part. It is possible to use built-in models in \marty\ such as the SM or the MSSM, or define a custom Lagrangian. The program we present next only requires the model to have fermions named \lstinline!"b"!, \lstinline!"s"! and \lstinline!"mu"!, and the photon \lstinline!"A"!.
\begin{framed}
\begin{lstlisting}
   PMSSM_Model model; // built-in pMSSM in MARTY
   // SM_Model model; // built-in SM in MARTY
\end{lstlisting}
\end{framed}

\paragraph{Settings for the calculation}
In order to extract the Wilson coefficients, \marty\ needs the convention-dependent global factor defined in the effective Hamiltonian (can be identified in equation~\ref{heff}) and the order of the external fermions in the operator because it is not uniquely defined. For the factor, it is first necessary to create its symbolic expression (using SM parameters in \lstinline!mty::sm_input::!), namely:
\begin{framed}
\begin{lstlisting}
   Expr factor = -pow_s(e_em, 4) * GetComplexConjugate(V_ts)
      * V_tb / (32 * CSL_PI*CSL_PI * M_W*M_W * s2_theta_W);
\end{lstlisting}
\end{framed}
Then, a \lstinline!FeynOptions! object can be created to contain the factor and the fermion order:\footnote{The fermions are defined from the order given to the \lstinline!computeWilsonCoefficients()! method. In the $b\to s\mu\mu$ process for example, the order $(1, 0, 2, 3)$ applied to the particles $(b,\bar{s},\bar{\mu},\mu)$ corresponds to an operator of the type $\left(\bar{s}b\right)\left(\bar{\mu}\mu\right)$.}
\begin{framed}
\begin{lstlisting}
   FeynOptions options;
   options.setWilsonOperatorCoefficient(factor);
   options.setFermionOrder({1, 0, 2, 3});
\end{lstlisting}
\end{framed}
More options can be set to customize the calculation, for example:
\begin{framed}
\begin{lstlisting}
   options.setTopology(Topology::Triangle); // Only triangles (3-point functions)
   options.addFilters(
        filter::disableParticle("A"), // No photon 
        filter::forceParticle("t") // Only top quark contributions
   );
\end{lstlisting}
\end{framed}
\paragraph{Extracting $C_{9(10)}$}
Once the options are set, the calculation of the $b\to s\mu\mu$ decay can be performed by \marty\ simply using:
\begin{framed}
\begin{lstlisting}
   auto wilsons = model.computeWilsonCoefficients(
         OneLoop, 
         {Incoming("b"), Outgoing("s"), Outgoing("mu"), Outgoing(AntiPart("mu"))},
         options // The options as last parameter
         );
\end{lstlisting}
\end{framed}
In order to extract the coefficients in the amplitude, \marty\ now has to identify operators in the result. These must be provided by the user but a simple interface exists to create any fermion currents,\footnote{It is also possible to define general effective operators in \marty\ based on a technique similar to the one used for the creation of Lagrangian terms.} in this case $d=6$ operators:
\begin{framed}
\begin{lstlisting}
   auto O9  = dimension6Operator(model, wilsons, DiracCoupling::VL, DiracCoupling::V);
   auto O10 = dimension6Operator(model, wilsons, DiracCoupling::VL, DiracCoupling::A);
\end{lstlisting}
\end{framed}
Using the fermion order defined in the options, it is possible to give the Dirac couplings corresponding to the two currents (Scalar \lstinline!(S)!, Pseudoscalar \lstinline!(P)!, Left \lstinline!(L)!, Right \lstinline!(R)!, Vector \lstinline!(V)!, Axial \lstinline!(A)!, Vector Left \lstinline!(VL)!, Vector Right \lstinline!(VR)!, Tensor \lstinline!(T)!, Tensor Axial \lstinline!(TA)!, Tensor Left \lstinline!(TL)! or Tensor Right \lstinline!(TR)!). Once $O_9$ and $O_{10}$ have been defined, the coefficients can be extracted:
\begin{framed}
\begin{lstlisting}
   Expr C9  = getWilsonCoefficient(wilsons, O9);
   Expr C10 = getWilsonCoefficient(wilsons, O10);
\end{lstlisting}
\end{framed}
Finally, \marty\ can generate the library containing the numerical functions evaluating $C_9$, $C_{10}$ and the tree-level spectrum generator for the model: 
\begin{framed}
\begin{lstlisting}
   Library lib("wilson_pmssm"); // Library called "wilson_pmssm"
   lib.generateSpectrum(model); // Provides the spectrum generator for the model
   lib.addFunction("C9_pmssm",  C9);
   lib.addFunction("C10_pmssm", C10);
   lib.build(); // Creates and builds the library automatically
\end{lstlisting}
\end{framed}
\paragraph{Extracting $C_7^{(\prime)}$}
For the extraction of $C_7^{(\prime)}$ in the $b\to s\gamma$ decay, the factor must be modified and the calculation is different:\footnote{The function is called \lstinline!chromoMagneticOperator()! because it works for general external bosons, including those with a non-trivial gauge representation such as the gluon for which an algebra generator has to be supplemented in the operator.}
\begin{framed}
\begin{lstlisting}
   auto wilsons = model.computeWilsonCoefficients(
         OneLoop,
         {Incoming("b"), Outgoing("s"), Outgoing("A")},
         options // The options as last parameter
         );
   auto O7  = chromoMagneticOperator(model, wilsons, DiracCoupling::R); // P_R
   auto O7p = chromoMagneticOperator(model, wilsons, DiracCoupling::L); // P_L
   Expr C7  = getWilsonCoefficient(wilsons, O7);
   Expr C7p = getWilsonCoefficient(wilsons, O7p);
\end{lstlisting}
\end{framed}

\section{Conclusion}

We presented a general procedure to extract $C_7$, $C_7^\prime$, $C_9$ and $C_{10}$ in any BSM model using \martyv. Such a procedure is very useful to obtain, in a fully automated way, the loop-level predictions of flavor observables in general BSM scenarios that is a very long and difficult task by hand. Furthermore, \marty\ is not limited to these Wilson coefficients and can also be used to extract coefficient of general $d\leq 6$ operators. Considering also the capability of \marty\ to calculate squared amplitudes at tree-level or one-loop for general BSM, this software program allows us to automate theoretical predictions at the loop level for general BSM scenarios.

\bibliographystyle{h-physrev5}  
\bibliography{biblio}

\end{document}